\documentclass[12pt,epsf]{article}

\usepackage{soul}

\def\man{\mathcal{M}}

\usepackage{graphicx,subfigure}
\usepackage{mathrsfs,hyperref}
\usepackage{amssymb,amsfonts}
\usepackage{amsmath,color}
\DeclareMathOperator{\arcsinh}{arcsinh}
\DeclareMathOperator{\arctanh}{arctanh}

\newcommand{\be}{\begin{equation}}

\newcommand{\ee}{\end{equation}}

\newcommand{\ba}{\begin{eqnarray}}

\newcommand{\ea}{\end{eqnarray}}

\begin{document}
\begin{titlepage}
\vspace{.5in}
\begin{flushright}
UdeM-GPP-TH-17-258
\end{flushright}
\vspace{0.5cm}

\begin{center}
{\Large\bf Tunneling decay of false vortices with gravitation }\\
\vspace{.4in}

  {$\rm{\acute{E}ric \,\, Dupuis}^{\P}$}\footnote{\it email: eric.dupuis.1@umontreal.ca},\,\,
  {$\rm{Yan \,\, Gobeil}^{\P\ddag}$}\footnote{\it email: yan.gobeil@mail.mcgill.ca},\,\,
  {$\rm{Bum-Hoon \,\, Lee}^{\S\dag}$}\footnote{\it email: bhl@sogang.ac.kr},\,\,
  {$\rm{Wonwoo \,\, Lee}^{\S}$}\footnote{\it email: warrior@sogang.ac.kr},\, \,
  {$\rm{Richard \,\, MacKenzie}^{\P}$}\footnote{\it email: richard.mackenzie@umontreal.ca} \\
  {$\rm{Manu\,\, B.\, Paranjape}^{\P}$}\footnote{\it email: paranj@lps.umontreal.ca},\, \,
  {$\rm{Urjit\,\, A.\, Yajnik}^{*}$}\footnote{\it email: yajnik@iitb.ac.in},\, \,
  {$\rm{Dong-han\,\, Yeom}^{\Re\natural}$}\footnote{\it email: innocent.yeom@gmail.com}  \\

{\small \P \it Groupe de Physique des Particules, D$\acute{e}$partement de physique, Universit$\acute{e}$ de Montr$\acute{e}$al,\\
               C. P. 6128, Succursale Centre-ville, Montr$\acute{e}$al, Qu$\acute{e}$bec, Canada H3C 3J7}\\
{\small \ddag \it Department of Physics, Ernest Rutherford Physics Building,
McGill University, 3600 rue University, Montr$\acute{e}$al, Qu$\acute{e}$bec, Canada H3A 2T8}\\
{\small \S \it Center for Quantum Spacetime, Sogang University, Seoul 121-742, Korea}\\
{\small \dag \it Department of Physics, Sogang University, Seoul 121-742, Korea}\\
{\small $*$ \it Department of Physics, Indian Institute of Technology Bombay, Mumbai 400076, India}\\
{\small $\Re$ \it Leung Center for Cosmology and Particle Astrophysics, National Taiwan University, Taipei 10617, Taiwan}\\
{\small $\natural$ \it Asia Pacific Center for Theoretical Physics, and Department of Physics, Pohang University of Science and Technology, Pohang 790-784, Korea}\\

\vspace{.5in}
\today
\end{center}

\begin{center}
{\large\bf Abstract}
\end{center}
\begin{center}
\begin{minipage}{4.75in}

{\small \,\,\,\,We study the effect of vortices on the tunneling decay of a symmetry-breaking false vacuum in three spacetime dimensions with gravity. The scenario considered is one in which the initial state, rather than being the homogeneous false vacuum, contains false vortices. The question addressed is whether, and, if so, under which circumstances, the presence of vortices has a significant catalyzing effect on vacuum decay.
After studying the existence and properties of vortices, we study their decay rate through quantum tunneling using a variety of techniques. In particular, for so-called thin-wall vortices we devise a one-parameter family of configurations allowing a quantum-mechanical calculation of tunneling. Also for thin-wall vortices, we employ the Israel junction conditions between the interior and exterior spacetimes. Matching these two spacetimes reveals a decay channel which results in an unstable, expanding vortex.
We find that the tunneling exponent for vortices, which is the dominant factor in the decay rate, is half that for Coleman-de Luccia bubbles. This implies that vortices are short-lived, making them cosmologically significant even for low vortex densities. In the limit of the vanishing gravitational constant we smoothly recover our earlier results for the decay of the false vortex in a model without gravity.}


\end{minipage}
\end{center}
\end{titlepage}



\newpage

\section{Introduction \label{sec1}}

The Abelian Higgs model in three spacetime dimensions has soliton solutions known as Nielsen-Olesen vortices \cite{Abrikosov:1956sx}. These objects have a localized magnetic flux proportional to their winding number; in four dimensions they correspond to magnetic flux tubes in type-II superconductors \cite{Nielsen:1973cs}, while in the cosmological context they could appear as cosmic strings \cite{Kibble:1976sj, Kibble:1980mv}, possibly playing a role in structure formation or having other observable effects \cite{Brandenberger:2013tr, Tashiro:2013xra}.

In the minimal model (with symmetry-breaking $\phi^4$ potential), vortices are built out of the true vacuum and are stable both classically and quantum mechanically. (Their classical stability is easy to see: the potential and scalar field gradient energies favor collapse, while the magnetic field energy favors expansion. The stable configuration is a compromise between these two antagonistic effects.)

In this paper, we wish to study vortices in non-minimal models. First, we consider a different potential where in addition to the symmetry-breaking vacuum there is a lower-energy symmetric vacuum; these are thus false and true vacua, respectively. Second, we include gravity. Depending on the details of the potential and the strength of the gravitational coupling, vortices can be rendered classically unstable. If classically stable, they will be metastable quantum-mechanically. Their lifetime can be quite short; if so, they could play an important role wherever they appear
(phase transitions in the early universe, for instance).

In earlier work \cite{Lee:2013ega}, a metastable analog of the vortex dubbed a false vortex was studied. It owes its name to the vacuum structure permitting its formation. Namely, the \textit{false} vortex is a topologically non-trivial solution built from a \textit{false} vacuum which corresponds to the spontaneously broken sector of a modified abelian Higgs model. The presence of a symmetry-restoring true vacuum for scalar field $\phi=0$ explains the vortex metastability: This lower-energy phase of the scalar field is contained within the vortex core and spoils protection from expansion. For this reason, the vortex can lead to interesting consequences for the cosmological history of this model.

We extend this work by considering the effect of gravity on the tunneling decay of false vortices in three spacetime dimensions. First of all, we must specify the zero of the potential energy, which becomes important in the presence of gravity. We choose the energy density in the symmetry-restoring true vacuum to be negative, while that in the false vacuum vanishes. This implies a modification of the spacetime. The exterior spacetime with a vortex confined within a finite radius is locally Minkowski with a conical defect \cite{Deser:1983tn}. The conical defect is the analog of the Schwarzschild mass parameter that is familiar in the $3+1$ dimensional context. Inside the vortex, the situation is more complex with a varying magnetic field, a negative energy density  and the scalar field's gradient energy.

Nevertheless, an analytical understanding is possible in the limit of a large topological winding number, which implies a large magnetic flux inside the vortex. In this case, the scalar field makes a sharp transition from the core where $\phi = 0$ to its asymptotic value; there is thus a thin wall separating the two vacua. Were it not for the magnetic flux, the interior of the vortex would then be exactly anti-de Sitter (AdS) spacetime. Instead, the exact solution of the $2+1$ dimensional Einstein-Maxwell equations with cylindrically symmetric, covariantly constant magnetic field in asymptotically AdS spacetime corresponds to the solution studied by Hirschmann and Welch \cite{Hirschmann:1995he}.

The case of thin-wall vortices is reminiscent of the thin-wall limit of false vacuum decay treated in the seminal papers of Coleman and collaborators \cite{Coleman:1977py, Callan:1977pt, Coleman:1980aw} and as in those works the thin-wall nature of the vortex allows for an approximate collective-coordinate-like treatment of vortex tunneling. We adopt a circular disc of Hirschmann-Welch spacetime \cite{Hirschmann:1995he} separated by a thin circular transition region (the wall), from Minkowski spacetime with a conical defect on the outside. To understand if the vortex can expand dynamically after tunneling and to understand the tunnelling process itself, we construct the Israel junction conditions \cite{Israel:1966rt}

The paper is organized as follows: In Section \ref{sec2} we set up the basic framework with the equations of motion and appropriate boundary conditions. We also present the numerical solutions of the vortices coupled to Einstein gravity. In Section \ref{sec3}, we specialize our study to thin-wall vortices, including constructing the junction conditions.  In Section \ref{sec4}, we compute tunneling rates for thin-wall vacuum decay and vortex disintegration. In Section \ref{sec5}, we summarize our results and discuss possible future applications.

\section{Vortex solutions \label{sec2}}

In this section, we derive the equations of motion for the gauge, scalar field, and metric functions and impose the appropriate boundary conditions. We present the numerical solutions for both thick- and thin-wall vortices. By varying the difference between the false and true vacuum states, we investigate how the existence of the solution can be affected by the strength of the gravitational constant.

\subsection{Setup and equations of motion}
\begin{figure}
\begin{center}
\includegraphics[width =3.2 in]{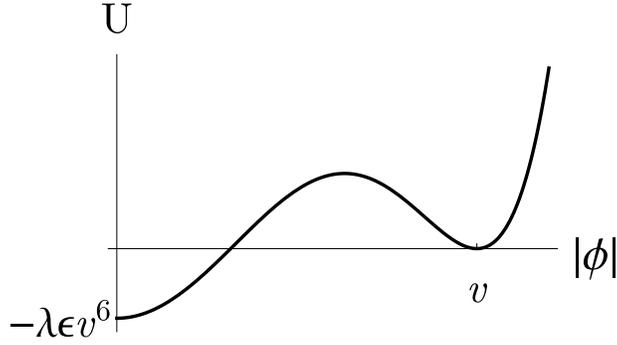}
\end{center}
\caption{Potential energy density.}
\label{fig:pot_phi6}
\end{figure}

We consider the action for Einstein gravity coupled to gauge and complex scalar fields:
\begin{equation}
S= \int_{\mathcal M} \sqrt{-g}\, d^3 x \left[ \frac{R}{2\kappa} - \frac{1}{16\pi}F_{\mu\nu}F^{\mu\nu} - (D_{\mu}\phi)^*(D^{\mu}\phi)-U(\phi^*\phi)
\right] + \int_{\partial \mathcal{M}} d^2x \sqrt{h} \frac{K}{\kappa} \,, \label{fvg-action}
\end{equation}
where $g=\det g_{\mu\nu}$, $\kappa \equiv 8\pi G$, $R$ denotes the curvature scalar of the spacetime $\mathcal M$, and $h$ is the determinant of the first fundamental form. $K$ is the trace of the second fundamental form of the boundary $\partial \mathcal{M}$ \cite{York:1972sj, Gibbons:1976ue}.
We adopt the sign conventions in Ref.\ \cite{Misner:1974qy}. The field strength tensor is $F_{\mu\nu} = \nabla_{\mu}A_{\nu} - \nabla_{\nu}A_{\mu}$, in which $A_{\mu}$ is the electromagnetic potential. The covariant derivative of a complex field is given by $D_{\mu}\phi = (\nabla_{\mu} + ie A_{\mu})\phi$, where $e$ is the coupling constant between gauge and complex scalar fields. The potential, shown in Fig.~\ref{fig:pot_phi6}, is given by
\begin{equation}
U(\phi^*\phi) = \lambda(|\phi|^2-\epsilon v^2)(|\phi|^2-v^2)^2\,, \label{potential}
\end{equation}
where $\lambda$ is the self-coupling constant of the scalar field. The value of $\epsilon$ determines the shape of the potential \cite{Lee:2013ega, Kumar:2010mv}. The curvature scalar has mass dimension $2$, $1/\kappa$ has mass dimension $1$, the fields $A_{\mu}$ and $\phi$, the charge all have mass dimension $1/2$, while the constants $\lambda$ and $\epsilon$ are dimensionless parameters. We are interested in the case where the false vacuum state is located at $\phi=v$ and the true vacuum state is located at $\phi=0$, therefore $0 < \epsilon < 1$. The geometry for the true vacuum state corresponds to AdS spacetime with an effective cosmological constant $\Lambda_{\rm eff}= \kappa U(0) = -\kappa \lambda \epsilon v^6$. In the remainder of this paper, computations will involve the absolute value of the cosmological constant, $\Lambda = |\Lambda_{\rm eff}| = \kappa \lambda \epsilon v^6$.

By varying the action, we obtain the Einstein equation:
\begin{equation}
R_{\mu\nu}-\frac{1}{2} R g_{\mu\nu} = \kappa T_{\mu\nu}\,, \label{Eequa}
\end{equation}
where the energy-momentum tensor has the form
\begin{eqnarray}
T_{\mu\nu}&=&\frac{1}{4\pi}(F_{\mu\alpha}F_{\nu}^{~\alpha} - \frac{1}{4}g_{\mu\nu} F_{\alpha\beta}F^{\alpha\beta}) +(D_{\mu}\phi)^*(D_{\nu}\phi)+ (D_{\mu}\phi)(D_{\nu}\phi)^* \nonumber \\
&-& g_{\mu\nu}[(D_{\alpha}\phi)^*(D^{\alpha}\phi)+ U(\phi^*\phi)]\,. \label{emten}
\end{eqnarray}

The gauge field and scalar field equations are, respectively,
\begin{equation}
\nabla_{\nu} F^{\nu}_{~\mu} = 4\pi[ i e (\phi \nabla_{\mu} \phi^* - \phi^* \nabla_{\mu}\phi ) + 2 e^2 A_{\mu} \phi^* \phi]\,, \label{gfequa}
\end{equation}
\begin{equation}
D_{\mu}D^{\mu} \phi = \frac{\partial U(\phi^* \phi)}{\partial \phi^*}\,. \label{sfequa}
\end{equation}

We take the metric ansatz as to be
\begin{equation}
ds^2= - A^2(t,r) dt^2 + B^{-2}(t,r) dr^2 + D^2(t,r) d\theta^2 \,,
\label{metric}
\end{equation}
where $A(t,r)$, $B(t,r)$, and $D(t,r)$ are unknown functions representing a rotationally invariant solution.

We now rewrite the equations in terms of dimensionless variables
\begin{equation}
\frac{\phi}{v} = \tilde{\phi},~ \frac{A_{\mu}}{v} = \tilde{A}_{\mu},~ \frac{e}{\sqrt{\lambda}v} =\tilde{e},~ \kappa v^2=\tilde{\kappa},~ r v^2\sqrt{\lambda}=\tilde{r}.
\label{dimensionless}
\end{equation}
In what follows, we use these dimensionless variables, suppressing the tildes for notational simplicity.

We look for solutions for $\phi$ and $A_{\mu}$ in the coordinates $(r$, $\theta)$. The field ansatz is
\begin{equation}
\phi(t,r, \theta) = f(t,r) e^{in\theta}, \;\;\;\;  A_{\mu}(t,r, \theta)=\left[0, 0, \frac{n(a(t,r)-1)}{e} \right] \,,
\label{ansatz}
\end{equation}
where $n$ is an integer, the winding number. Using this ansatz, different terms appearing in the action can be reduced as follows:
\begin{equation}
D_{\mu}\phi = (\nabla_{\mu} + ie A_{\mu})\phi = (\dot{f}, f', i n a f) e^{i n\theta} \,, \nonumber
\end{equation}

\begin{equation}
(D_{\mu}\phi)^*(D^{\mu}\phi) = \left(-\frac{\dot{f}^2}{A^2}+ f'^2B^2 + \frac{n^2f^2a^2}{D^2}  \right) \,,
\end{equation}

\begin{displaymath}
F_{\mu\nu} = \frac{n}{e}
\left(\begin{array}{ccc}
0&0&\dot{a}\\
0&0&a'\\
-\dot{a}&-a'&0
\end{array}\right) \,,
\end{displaymath}

\begin{equation}
F_{\mu\nu}F^{\mu\nu} = \frac{2n^2}{e^2 D^2}\left(-\frac{\dot{a}^2}{A^2} + a'^2B^2  \right) \,,
\end{equation}
where the dot and the prime denote differentiation with respect to $t$ and $r$, respectively.

Using these results, the equations of motion are written out as a function of $f, a, A, B ,D$ fields. First, the $(tt)$, $(tr)$, $(rr)$, and $(\theta\theta)$ components of the Einstein equations are
\begin{eqnarray}
&&-\frac{A^2 B^2 (B'D'+BD'')+\dot{B}\dot{D}}{BD} \nonumber \\
&&=\kappa \left[ \frac{n^2\dot{a}^2}{8\pi e^2 D^2} + \dot{f}^2 + \frac{A^2 n^2}{8\pi e^2 D^2} a'^2 B^2 + A^2 \left(f'^2B^2 + \frac{n^2 a^2 f^2}{ D^2} + U \right) \right]\,,  \label{eqtr1}
\end{eqnarray}
\begin{equation}
-\frac{1}{D}\left(\frac{D'\dot{B}}{B} - \frac{A'\dot{D}}{A} + \frac{\dot{D}'}{A}\right) =\kappa \left[ \frac{n^2\dot{a}a'}{4\pi e^2 D^2} + 2\dot{f}f' \right]\,, \label{eqtr2}
\end{equation}
\begin{equation}
\frac{A^2B^2A'D'+\dot{A}\dot{D}-A\ddot{D}}{A^3B^2D} = \kappa \left[ \frac{1}{A^2B^2}\left(\dot{f}^2 +\frac{n^2 \dot{a}^2}{8\pi e^2 D^2}\right) + \frac{n^2 a'^2}{8\pi e^2 D^2} + f'^2 - \frac{n^2 f^2a^2}{B^2D^2} -\frac{U}{B^2} \right]\,, \label{eqtr3}
\end{equation}
\begin{eqnarray}
&&\frac{D^2[A^2B^3(A'B'+BA'')-B\dot{A}\dot{B}+A(-2\dot{B}^2 +B\ddot{B})]}{A^3 B^2} \nonumber \\
&&= \kappa \left[ \frac{1}{A^2}\left(\dot{f}^2D^2 - \frac{n^2\dot{a}^2}{8\pi e^2} \right) + \frac{n^2 a'^2 B^2}{8\pi e^2 } + n^2a^2f^2 - f'^2B^2D^2 -UD^2 \right]\,. \label{eqtr4}
\end{eqnarray}
As for the matter, the scalar field equation has the form
\begin{equation}
\frac{1}{A^2}\left[-\ddot{f} +\left(\frac{\dot{A}}{A} + \frac{\dot{B}}{B} -\frac{\dot{D}}{D} \right) \dot{f} \right] +  B^2 \left[ f'' + \left(\frac{A'}{A} +\frac{B'}{B} + \frac{D'}{D} \right) f'\right] - \frac{n^2 a^2}{D^2} f = \frac{dU}{df}\,, \label{eqtr5}
\end{equation}
while that of the gauge field is
\begin{equation}
\frac{1}{A^2}\left[-\ddot{a} +\left(\frac{\dot{A}}{A} + \frac{\dot{B}}{B} +\frac{\dot{D}}{D} \right) \dot{a} \right] +  B^2 \left[ a'' + \left(\frac{A'}{A} +\frac{B'}{B} - \frac{D'}{D} \right) a'\right] = 8\pi e^2 f^2 a\,. \label{eqtr6}
\end{equation}

We wish to find the static configuration of the vortex with gravity. Even in this case, since exact analytic solutions have not been found even without gravity, we solve the equations numerically. Because the metric functions only depend on $r$, we are free to choose a gauge in which $B(r) = 1$ everywhere. This was the approach used in Refs.\ \cite{Garfinkle:1985hr, LagunaCastillo:1987cs}.
We simultaneously solve the coupled Einstein, gauge, and scalar field equations with the following boundary conditions:
\begin{eqnarray}
f(r) \rightarrow 0,  \quad a(r) \rightarrow 1, \quad A'(r) \rightarrow 0, \quad D(r) \rightarrow 0, \quad D'(r) \rightarrow 1, \quad &&{\rm as} \quad r \rightarrow 0\,,\label{boundary1} \\
f(r) \rightarrow 1,  \quad a(r) \rightarrow 0, \quad A(r) \rightarrow 1, \quad &&{\rm as} \quad r \rightarrow \infty \,.  \label{boundary2}
\end{eqnarray}
The first conditions arise from the requirement that the solution be nonsingular at the origin and the second conditions are required for a solution of finite energy. In the absence of gravitation, the behavior for small $r$ can be analyzed by linearizing the matter field equations, so that $f(r) \sim r^n$ and $ a(r) \sim r^2$ as $r \rightarrow 0$. For large $r$, we write $f(r) = 1- \xi(r)$ and linearize in $\xi(r)$ and $a(r)$, resulting in modified Bessel equations \cite{Rubakov:2002fi}. As for the field $A(r) = \sqrt{g_{tt}}$, multiple boundary conditions are possible with simple rescaling of time since the metric is time independent. $A(\infty)$ is fixed to $1$ so that time is properly normalized for asymptotic observers. In any case, this is simply a matter of time rescaling, as the metric is time-independent.

\subsection{Numerical solutions \label{sec2-2}}

We numerically solve the coupled equations of the gauge, scalar field, and gravity simultaneously. For the static configuration we solve   (\ref{eqtr1} - \ref{eqtr6}), the terms without time derivatives, with the boundary conditions (\ref{boundary1}) and (\ref{boundary2}).

\begin{figure}[ht!]
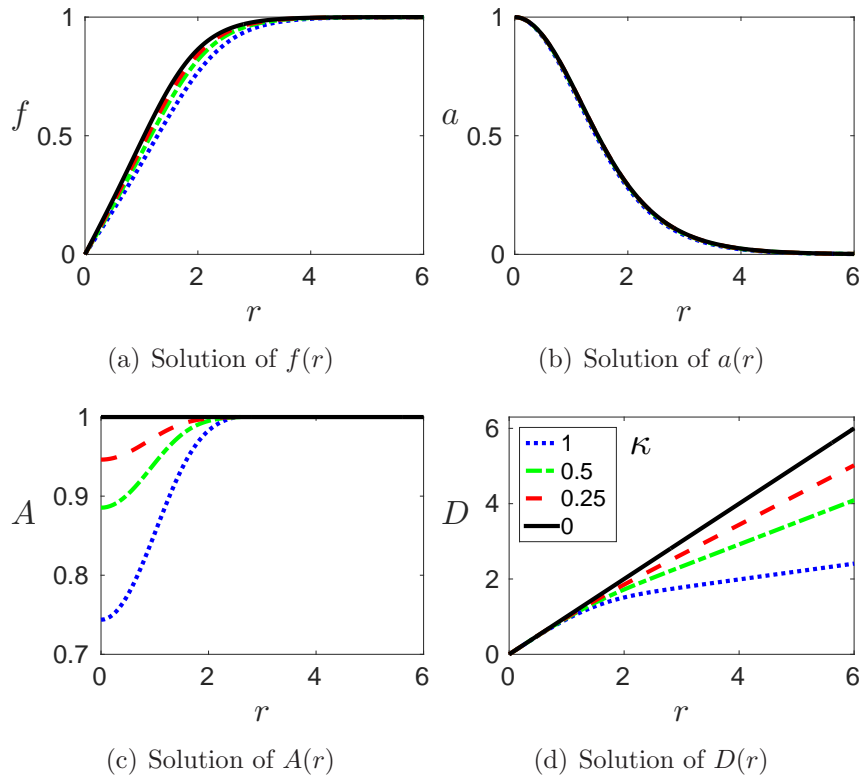

\begin{center}
\subfigure[Solution of $f(r)$]
{\includegraphics[width =2.2 in]{scthickf.eps}}
\subfigure[Solution of $a(r)$]
{\includegraphics[width =2.2 in]{scthicka-.eps}}\\
\subfigure[Solution of $A(r)$]
{\includegraphics[width =2.2 in]{scthickA.eps}}
\subfigure[Solution of $D(r)$]
{\includegraphics[width =2.2 in]{scthickD.eps}}
\end{center}
\caption{Thick-wall solutions of $f(r)$, $a(r)$, $A(r)$ and $D(r)$ respectively. $n~=~e /\sqrt{4 \pi}~=~1$,  $\epsilon~=~0.1$. The line patterns (solid line, dashed line, dash-dotted line, dotted line) correspond to $\kappa = (0.00,0.25,0.50,1.00)$, respectively.}
\label{Numerical-solutions-k-1}
\end{figure}

\begin{figure}[ht!]
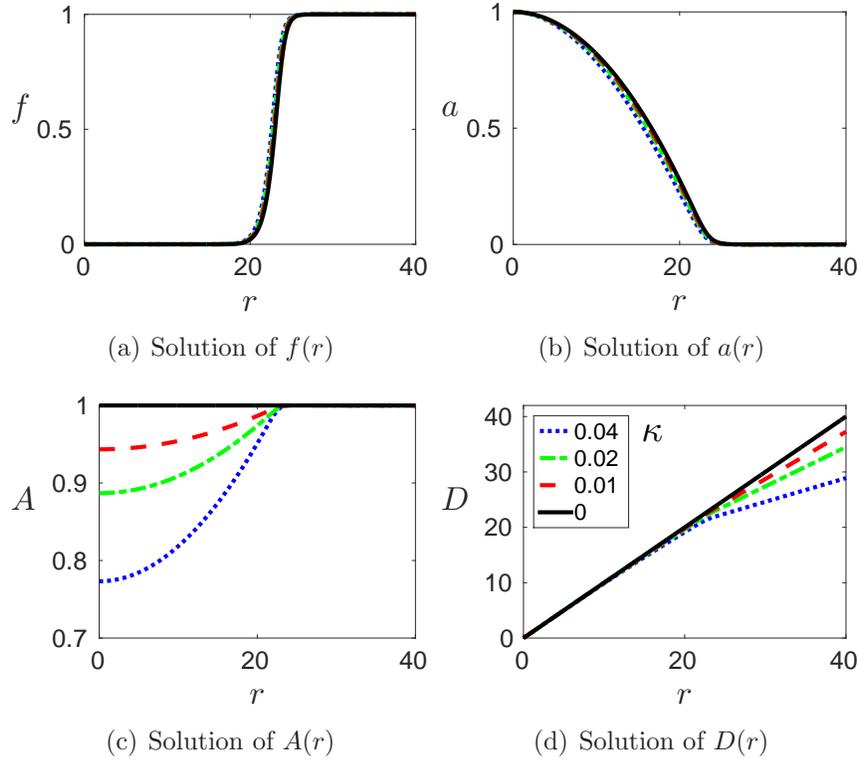

\begin{center}
\subfigure[Solution of $f(r)$]
{\includegraphics[width =2.2 in]{scthinf.eps}}
\subfigure[Solution of $a(r)$]
{\includegraphics[width =2.2 in]{scthina-.eps}}\\
\subfigure[Solution of $A(r)$]
{\includegraphics[width =2.2 in]{scthinA.eps}}
\subfigure[Solution of $D(r)$]
{\includegraphics[width =2.2 in]{scthinD.eps}}
\end{center}
\caption{Thin-wall solutions of $f(r)$, $a(r)$, $A(r)$, and $D(r)$  respectively. $n~=~50,\; e/\sqrt{4 \pi}~=~1$, $\epsilon~=~0.005$. The line patterns (solid line, dashed line, dash-dotted line, dotted line) correspond to $\kappa = (0.00,0.01,0.02,0.04)$, respectively.}
\label{fig:thin}
\end{figure}

Numerical solutions for $f(r)$, $a(r)$, $A(r)$, $B(r)$, and $D(r)$ are shown in Figs.\ \ref{Numerical-solutions-k-1} and \ref{fig:thin}. The profile of $f(r)$ is used to categorize two general types of vortex solutions. For small value of $n$, near $1$, solutions are dubbed ``thick": The transition from true vacuum to false vacuum shown by $f(r)$ happens on a relatively large scale. Such solutions are shown for $n=1$ in Fig.~\ref{Numerical-solutions-k-1}. On the other hand, for $n \gg 1$ the solutions transit rapidly giving the thin-wall behavior described earlier, as can be seen from Fig.~\ref{fig:thin} for $n=50$. In both cases, solutions obtained with different values of $\kappa$ are plotted. The vortex profile for $\kappa=0$ is in accordance with the previous study \cite{Lee:2013ega}. As gravity is added, matter fields do not change much compared to metric fields. For a conical spacetime, it is expected that  $D(r) = \sqrt{g_{\theta \theta}} = \left (1-4 G \mu \right) r$ where $\mu$ is the energy of the localized source \cite{Vilenkin:1984ib}. This behavior is indeed observed: Outside the vortex, $D(r)$ is linear, with a slope which decreases as $\kappa$ increases. As for $A(r)$, it departs further from $1$ at the origin for increasing values of $\kappa$. The actual behavior of $A(r)$, getting smaller or bigger compared to its flat spacetime value of $1$, should depend on the scalar field/gauge field mass ratio, as was observed in \cite{LagunaCastillo:1987cs}. We also noted that there was a maximal value $\kappa$ beyond which we could not find solutions anymore.

\begin{figure}[ht!]
\begin{center}
\includegraphics[width =3.5 in]{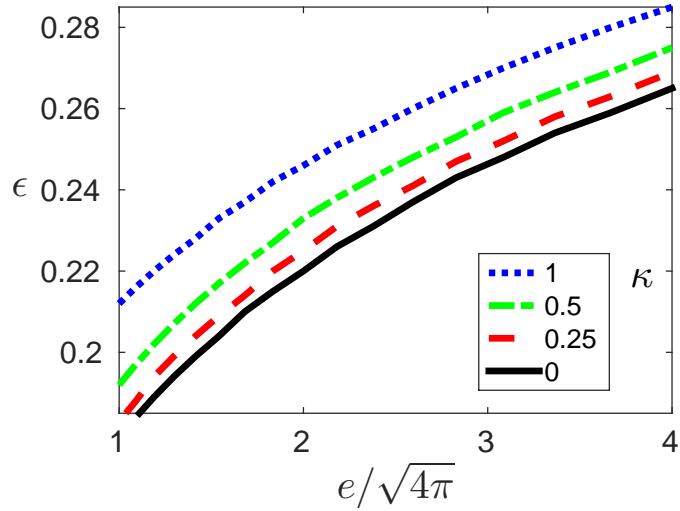}
\end{center}
\caption{The line patterns (solid line, dashed line, dash-dotted line, dotted line) represent the critical value $\epsilon_c$ as a function of $e$ for $\kappa = (0.00,0.25,0.50,1.00)$, respectively. For $\epsilon \geq \epsilon_c$, no metastable vortex solutions were found.}
\label{fig:eps_c}
\end{figure}

In the previous paper \cite{Lee:2013ega}, it was shown that the difference between the false and true vacuum state, $\epsilon$, is crucial for the metastability of the vortices. They become unstable above a certain critical value $\epsilon_c$. This behavior defines a region in parameter space in which vortex solutions cannot be found.  Fig.~\ref{fig:eps_c} shows a scan of the parameter space for $n=1$ and different values of $\kappa$. As $\kappa$ gets bigger, the allowed region for the formation of metastable vortices ($\epsilon < \epsilon_c$) is also expanded. The equivalent behavior for thin-wall solutions is studied in detail in the next section, in which we formulate proper junction conditions at the surface of the vortex.

\section{Junction conditions for the thin-wall vortex \label{sec3}}

The existence of static vortex solutions with gravity has been established. We now investigate more closely the metastability of thin-wall vortices. For this purpose, we employ the Israel junction conditions \cite{Israel:1966rt} to understand if the vortex can tunnel through a certain potential barrier and expand dynamically. We thus consider a thin wall partitioning bulk spacetime into two distinct three-dimensional manifolds, $\cal{M}^+$ and $\cal{M}^-$, with boundaries $\Sigma^+$ and $\Sigma^-$ for the inside $(-)$ and outside $(+)$ of the vortex, respectively. To obtain the single glued manifold $\cal{M}=\cal{M}^+ \bigcup \cal{M}^-$, we demand that the boundaries are identified as follows:
\begin{equation}
\Sigma^+ = \Sigma^- = \Sigma \,,
\end{equation}
where the thin-wall boundary $\Sigma$ is a timelike hypersurface with unit normal $n^{\mu}$.

The bulk spacetime geometry, as eq.\ (\ref{metric}), is described by the metric
\begin{equation}
ds^2_{(\pm)} = -A^2_{\pm}(r) dt^2 + B^{-2}_{\pm}(r)dr^2 + D^2_{\pm}(r) d\theta^2 \,. \label{eq:1stmetric}
\end{equation}

The energy-momentum tensor $T^{\mu\nu}$ has a singular component on the wall
\begin{equation}
T^{\mu\nu} = S^{\mu\nu}\delta(\eta) + {\rm regular~term} \,,
\end{equation}
where $S^{\mu\nu}(x^i , \eta=\bar\eta)$ is the surface stress-energy tensor of the wall
\begin{equation}
S_{\mu\nu} = \lim_{\epsilon \rightarrow 0} \int^{\bar\eta +
\epsilon}_{\bar\eta - \epsilon} T_{\mu\nu} d\eta \,,
\end{equation}
where $\delta \ll \bar{\eta}$. The extrinsic curvature has only two components, $k^{\theta}_{\theta}$ and
$k^{\tau}_{\tau}$. The form of the stress-energy tensor on the wall is obtained using the covariant conservation.

We introduce the Gaussian normal coordinate system near the wall
\begin{equation}
ds^2 = - d\tau^2 + d\eta^2 + {\bar R}^2(\tau, \eta) d\theta^2\,,
\end{equation}
where $g_{\tau\tau}=-1$ and ${\bar R}(\tau, \bar{\eta})=R(\tau)$.
It must agree with the coordinate $R$ on both sides of the junction. In this coordinate system, the induced metric
on the hypersurface is
\begin{equation}
ds^2_{(\Sigma)} = -d\tau^2 + R(\tau)^2 d\theta^2 \,,  \label{hyperinmetric}
\end{equation}
where $\tau$ is the proper  time measured on the wall and $R(\tau)$ is the proper radius of
$\Sigma$. Given the metric defined in eq.\ (\ref{eq:1stmetric}), the following relation is satisfied
\begin{equation}
-d\tau^2 = -A^2 dt^2 + B^{-2}dr^2 ~~~\Rightarrow~~~(-A^2 \dot{t}^2 + B^{-2} \dot{r}^2 =-1) \,,
\end{equation}
where $\cdot$ denotes the differentiation with respect to $\tau$ in this section.

The induced metric of the hypersurface is given by
\begin{equation}
h_{ab} = g_{\alpha\beta} e^{\alpha}_{a}e^{\beta}_{b} \,,
\end{equation}
where the tangent vectors are
\begin{equation}
e^{\alpha}_{\tau} = (\dot{t}, \dot{r}, 0), ~~~~ e^{\alpha}_{\theta} = (0,0, r D^{-1})\,.
\end{equation}
The three-velocity of any point on the wall is
\begin{equation}
u^{\alpha} = (\dot{t}, \dot{r}, 0), ~~~~ u_{\alpha} = (-A^2 \dot{t},B^{-2} \dot{r}, 0)\,,
\end{equation}
which satisfies the relation $u_{\alpha}u^{\alpha}=-1$. The normal vectors are
\begin{equation}
n^{\alpha} = B^{-1}A(A^{-2}\dot{r}, B^2 \dot{t}, 0), ~~~~ n_{\alpha} = B^{-1}A(-\dot{r}, \dot{t}, 0) \,,
\end{equation}
where we take the factor $B^{-1}A$ to normalize the vectors, so that $n_{\alpha}n^{\alpha}=1$.
The extrinsic curvature then becomes
\begin{equation}
k^{\theta}_{\theta} =  \partial_{\theta} n^{\theta} + \Gamma^{\theta}_{\theta\mu}n^{\mu} = \frac{D'}{D} \sqrt{B^2+\dot{r}^2} \,,
\end{equation}
where $\Gamma^{\theta}_{\theta r}=\frac{D'}{D}$.

The relevant junction condition is given by
\begin{equation}
k^{\theta}_{\theta}(\rm inside) - k^{\theta}_{\theta}(\rm outside) = \kappa\sigma \,,
\end{equation}
where $\sigma$ is the surface tension on the wall. Only the scalar field contributes to the tension. We ignore the contribution from the negligible magnetic flux on the wall.

\subsection{The junction equation \label{sec3-1}}

We take the outside geometry to be flat Minkowski spacetime minus a wedge, the deficit angle parameter $\Delta$:
\begin{equation}
ds^2_{(+)} = - dt^2 + dr^2 + (1-\Delta)r^2 d\theta^2\,. \label{out-min}
\end{equation}
In the vortex core, we employ the geometry as the magnetic solution in AdS spacetime \cite{Hirschmann:1995he}
\begin{equation}
ds^2_{(-)} = -N^2(r) dt^2 + \frac{L dr^2}{N^2(r)J(r)}  + \frac{r^2 J(r)}{L} d\theta^2\,,  \label{out-ads}
\end{equation}
where
\begin{equation}
N^2(r)\equiv(1+ L\Lambda r^2),\quad J(r)\equiv [1+(\kappa Q^2_m/(8 \pi Lr^2)) \ln(1+L\Lambda r^2)],\quad L \equiv (1+\Lambda \kappa Q^2_m / (8 \pi)).
\label{eq:defs}
\end{equation}
Here, $Q_m$ is related to the amplitude of the magnetic field flux. The peculiar notation is due to its association to a magnetic charge in \cite{Hirschmann:1995he}; this is somewhat misleading, as the magnetic field sourced by $Q_m$ is not oriented radially. Nevertheless, we maintain this notation for consistency.  $\Lambda$ is the absolute value of the cosmological constant, $\Lambda = |\Lambda_{\rm eff}| = \kappa \epsilon$. The metric presented here is a bit different from the original formulation of this spacetime in \cite{Hirschmann:1995he}: Factors of $L$ in \eqref{out-ads} appear after a rescaling of the variables that ensures that $\lim_{r \to 0} g_{\theta \theta} = r^2$, therefore avoiding the conical singularity at the origin. This geometry corresponds to a one-parameter family of solutions with a magnetic flux in three-dimensional AdS spacetime. For $Q_m=0$, the metric reduces to AdS spacetime. The magnetic field measured in an orthonormal basis is given by ${\mathcal B} =  Q_m \Lambda / \sqrt{(1+ L \Lambda r^2)}$. The field is maximal at the origin and decreases monotonically until the boundary.

We change the metric into the following since the two geometries do not have the same circumferential radius. After getting the equation with the effective potential, we will return to the original coordinate system. The outside geometry takes the form
\begin{equation}
ds^2_{(+)} = - dt^2 + \frac{d{\tilde r}^2}{(1-\Delta)}+ \tilde{r}^2 d\theta^2\,, \label{out-min}
\end{equation}
and the inside geometry takes the form
\begin{equation}
ds^2_{(-)} = -F^2(\tilde{r}) dt^2 + \frac{(dr/d{\tilde r})^2}{E^{2}(\tilde{r})} d\tilde{r}^2 + \tilde{r}^2 d\theta^2\,.  \label{out-ads2}
\end{equation}
Then we can make the junction condition determining the motion of the thin wall which is located at position $r=R$ (or $\tilde{R} =\tilde{r}|_{r=R}$)
\begin{eqnarray}
&& k^{\theta}_{\theta}(\rm inside) - k^{\theta}_{\theta}(\rm outside) =  \\ \nonumber
&& \sqrt{E^2(\tilde{R})(d\tilde{r}/dr)^2\Big|_{r=R}+\dot{\tilde{R}}^2} -  \sqrt{(1-\Delta)+\dot{\tilde{R}}^2} =   \kappa \sigma \tilde{R}\,.
\end{eqnarray}
After squaring twice, we obtain
\begin{equation}
\frac{1}{2} \dot{\tilde{R}}^2 + V_{\rm eff}(\tilde{R}) = 0\,, \label{eq:effpot}
\end{equation}
where the effective potential takes the form
\begin{equation}
V_{\rm eff}(\tilde{R})= \frac{(1-\Delta)}{2} - \frac{(E^2(d\tilde{R}/dR)^2-(1-\Delta) - \kappa^2\sigma^2\tilde{R}^2)^2}{8\kappa^2\sigma^2\tilde{R}^2} \,.
\end{equation}
We return to the original coordinate system by using the relation
\begin{equation}
d\tilde{R}=\frac{\sqrt{H(R)}dR}{\sqrt{L(R)}\sqrt{J(R)}}, \quad H(R)\equiv \left(1+\frac{\Lambda \kappa Q^2_m / (8 \pi)}{1+L\Lambda R^2} \right)^2 \,. \label{backcosyre}
\end{equation}
The reason why $L$ is expressed as a function of $R$ shall be explained shortly. First, eq.\ (\ref{eq:effpot}) can then expressed in terms of $R$
\begin{equation}
\frac{1}{2} \dot{R}^2 + V_{\rm eff}(R) = 0\,, \label{junceq}
\end{equation}
where the effective potential turns out to be
\begin{equation}
V_{\rm eff}(R)=\frac{L(R) J(R)}{H(R)}
\times \left\{\frac{(1-\Delta)}{2} - \frac{\left[ \frac{N^2(R)H(R)}{L^2(R)} -(1-\Delta) - \frac{\kappa^2\sigma^2 R^2 J(R)}{L(R)} \right]^2}{8\kappa^2\sigma^2 r^2 J(R)/L(R)} \right\} \,. \label{efpotential}
\end{equation}
Now, $L(r)$ defined in \ (\ref{eq:defs}) depends on $Q_m$ which in turn is related to $R$. The magnetic flux $\Phi$ is a constant because of its topological origin; as the vortex radius may change by quantum fluctuations, the following relation between $Q_m$ and $R$ holds:
\begin{equation}
\Phi \equiv \int^{R}_{0} \int^{2\pi}_{0} d^2 x \sqrt{\det g_{ij}}{\mathcal B} = \frac{Q_m \pi}{L} \ln (1+L\Lambda R^2) \,. \label{totalflux}
\end{equation}
This transcendental equation cannot be solved exactly for $Q_m$. Making an expansion in $\kappa$, the first-order quantity is $Q_m \simeq \Phi /\pi \Lambda R^2$. $Q_m$ should then be expressed as a function of $Q_m(\Phi, \epsilon, \kappa, R)$ to replace $L(R)$ in eq.\ (\ref{efpotential}).

\begin{figure}
\begin{center}
\subfigure[Potential turning points $R_0$ and $R_1$ for $\kappa~=~0$, $\epsilon=0.005$. \label{fig:eff_pot_a}]
{\includegraphics[height=4.5 cm]{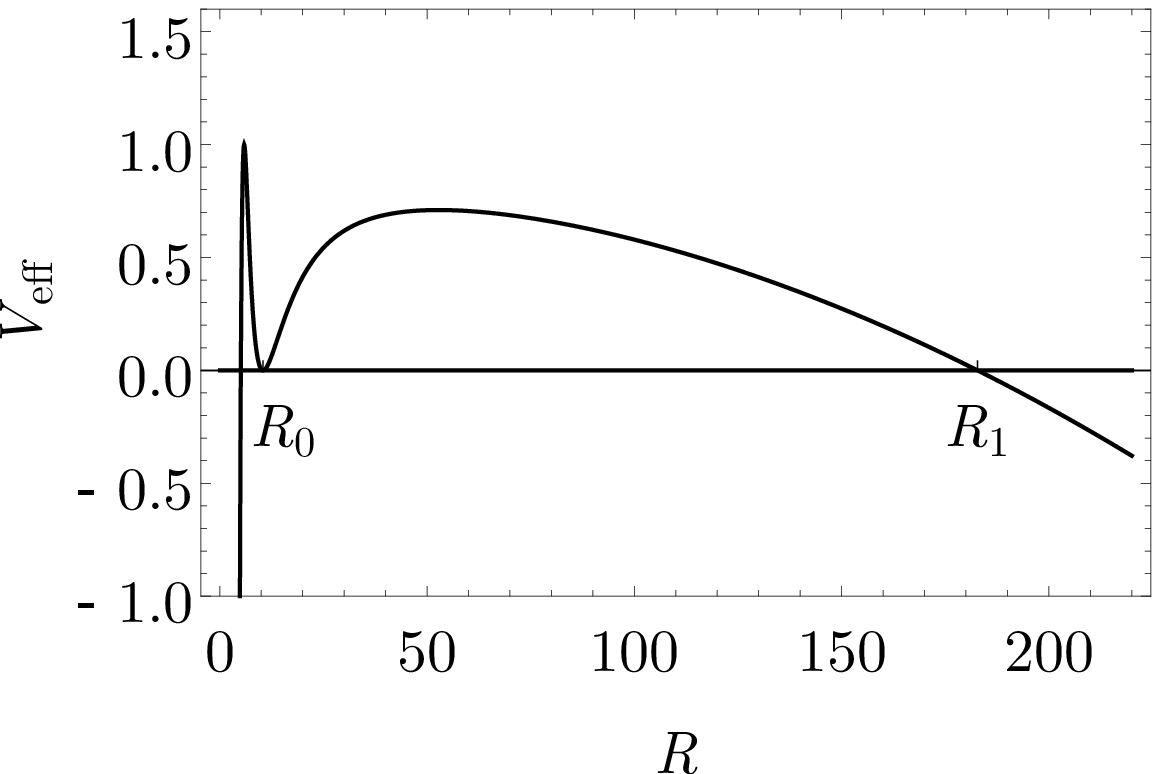}}
\;\;
\subfigure[The line patterns (solid line, dashed line, dash-dotted line, dotted line) respectively represent $\kappa = (0.0,0.5,1.0,2.5) \kappa_c$ and $\epsilon = 0.005$. The critical value of $\kappa$ is $\kappa_c \equiv \epsilon/\sigma^2$. \label{fig:eff_pot_b}]
{\includegraphics[height = 4.5 cm]{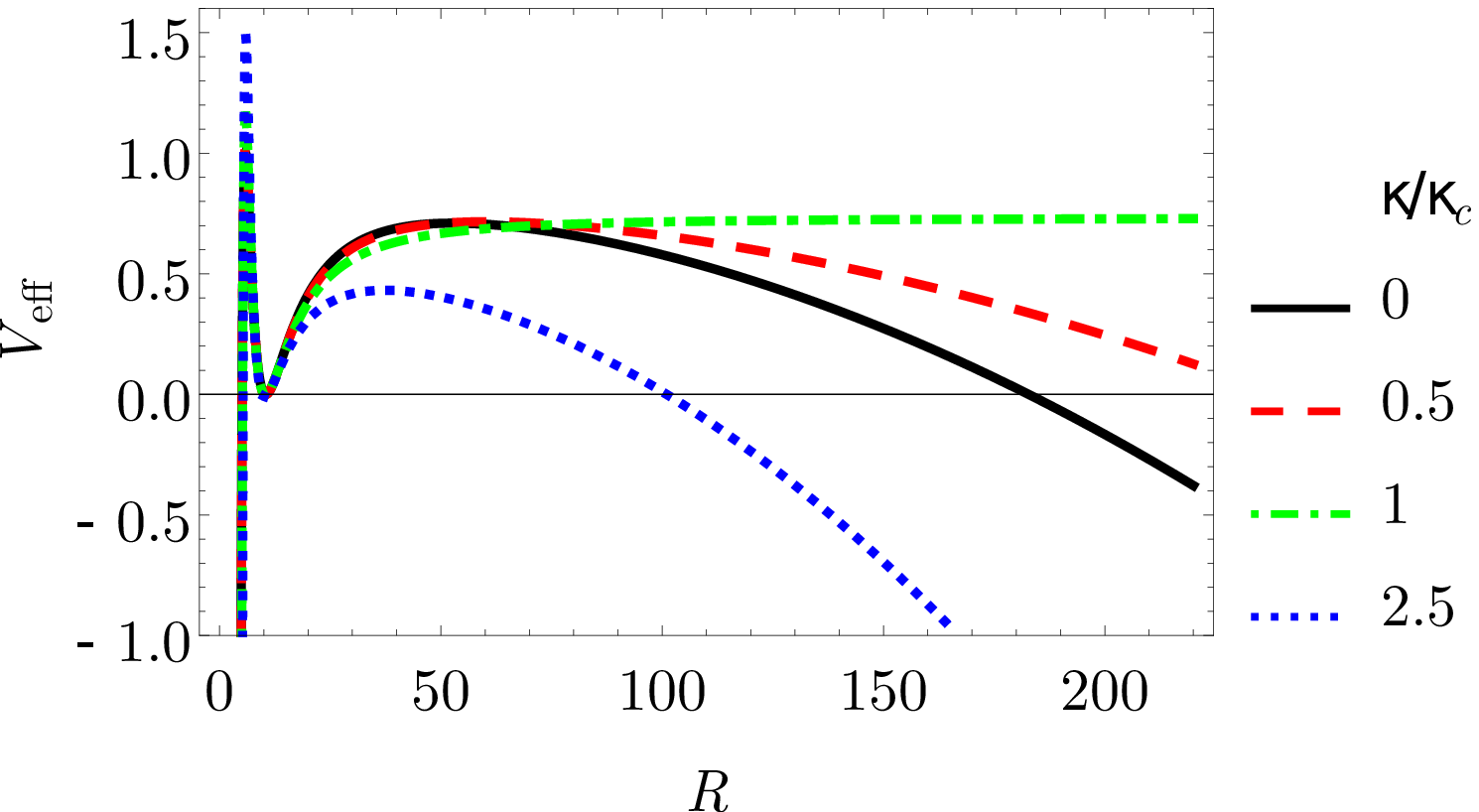}}
\end{center}
\caption{Effective potential for $\Phi=100$, $\epsilon=0.005$ and several values of the gravitational coupling constant $\kappa$.}
\label{fig:eff_pot}
\end{figure}

Fig.~\ref{fig:eff_pot_a} shows the shape of the effective potential with parameters $\Phi=100$, $\epsilon=0.005$ and $\kappa=0$ (as for $\Delta$ and $\sigma$, we show in the next section that they can be determined from the other parameters).  The main feature of this potential is the presence of an energy-vanishing minimum, which is located at $R_0$ in Fig.~\ref{fig:eff_pot_a}. According to Eq.\ (\ref{junceq}), this corresponds to a classically stable vortex. An energy barrier prevents the vortex from classically expanding to the escape radius $R_1$ from which it would explode; the vortex is quantum mechanically metastable. This feature of the system was observed in flat spacetime in \cite{Lee:2013ega} using different methods.

We study how this effective potential is affected by gravity. Fig.~\ref{fig:eff_pot_b} shows how the shape of the potential changes as the gravitational coupling constant is increased. Values of $\kappa$ are presented as a fraction of $\kappa_c = \epsilon/\sigma^2$, a critical value of $\kappa$ for which the vortex becomes completely stable. In the present case, we observe that the tunneling barrier of the effective potential gets bigger as $\kappa$ is increased towards $\kappa_c$. The opposite behaviour is seen for $\kappa > \kappa_c$; the barrier becomes weaker as $\kappa$ continues to increase. Eventually the barrier vanishes and no more thin-wall vortices can be formed. This agrees nicely with the absence of numerical classical solutions noted in Section \ref{sec2-2}. We note that the vortex can also shrink through quantum fluctuations and collapse. However, using the effective potential to compute the probability of collapsing
would be a stretch since the thin-wall approximation is only valid for large values of $R$. We concentrate  on the expansion metastability as we wish to compare this process to vacuum decay.

\subsection{The conical defect \label{sec3-2}}

As mentioned above, the vortex creates a conical defect (mass defect) in the angular coordinate. We examine the deficit angle in terms of the energy of the vortex configuration.

The effective potential (\ref{efpotential}) contains five parameters. Three of them, $\Phi$, $\epsilon$ and $\kappa$, are inputs. Parameters $\Delta$, the deficit angle parameter, and $\sigma$, the surface energy density on the wall, are determined by the first three parameters.  We will show the relation between them in the remainder of this section. Dependencies on $R$ are not shown explicitly to improve readability.

We first find $\Delta$ using eq.\ (\ref{junceq}) in the same way we found $V_{\rm eff}$
\begin{eqnarray}
\Delta &=& 1+ 2\kappa\sigma \tilde{R}(d\tilde{R}/dR) \sqrt{E^2+\dot{R}^2} -E^2(d\tilde{R}/dR)^2- \kappa^2\sigma^2 \tilde{R}^2 \nonumber \\
&=& 1+ 2\kappa\sigma R \frac{\sqrt{H}}{L} \sqrt{\frac{N^2J}{L}+ \dot{R}^2} - \frac{N^2 H}{L^2} - \kappa^2\sigma^2 R^2 \frac{J}{L}\,.
\end{eqnarray}
Expanding everything to first order in $\kappa$, we find
\begin{eqnarray}
\frac{\sqrt{H}}{L} &\simeq& 1- \frac{\kappa \Phi^2}{8 \pi^3 R^2} + {\cal O}(\kappa^2)\,, \\
\frac{N^2J}{L} &\simeq& 1+ \kappa\epsilon R^2 + \frac{\kappa \Phi^2}{16\pi^3 R^2}+ {\cal O}(\kappa^2)\,, \\
\frac{N^2 H}{L^2} &\simeq& 1+ \kappa\epsilon R^2 - \frac{\kappa \Phi^2}{8 \pi^3
 R^2}+ {\cal O}(\kappa^2)\,.
\end{eqnarray}
Using this, $\Delta$ can be approximated by
\begin{equation}
\Delta \equiv 8G\mu \simeq 8G \left(\frac{\Phi^2}{8\pi^2 R^2} + 2\pi\sigma R \sqrt{1+\dot{R}^2 } - \pi \epsilon  R^2 \right) + {\cal O}(G^2) \,, \label{deltaconide}
\end{equation}
where $\mu$ is the energy of the vortex configuration. As the conical defect should be conserved throughout the tunneling process due to energy conservation, we can simply evaluate it at the static vortex. Moreover, this energy of the static vortex should be minimized:
\begin{equation}
\frac{d\Delta|_{\dot{R}=0}}{dR}\Big|_{R=0}  = 0 \,.
\end{equation}
Neglecting terms of order $\epsilon$, one finds
\begin{equation}
r_0 = \left(\frac{\Phi^2}{8\pi^3\sigma}\right)^{1/3}, \quad \Delta_0 =8G\left(3\pi\sigma r_0 \right) \,.
\end{equation}

In the non-relativistic limit $\dot{R} \ll 1$, the energy of the vortex is
\begin{equation}
\mu \simeq \frac{\Phi^2}{8\pi^2 R^2} + 2\pi\sigma R \left(1+ \frac{1}{2}\dot{R}^2 \right) -  \pi \epsilon R^2 \,,
\end{equation}
where the quantized magnetic flux is given by $\Phi^2/4\pi=4\pi^2 n^2/e^2$. The factor $4\pi$ in the denominator is due to the convention of the field strength tensor in the action. This is almost the energy obtained in flat spacetime in \cite{Lee:2013ega} except for a missing kinetic term in the electromagnetic contribution. The reason for this discrepancy is that the interior metric written in  (\ref{out-ads}) was intended to be static. As we let spacetime change with varying radius, the functional form should also be changed. As we will see later, the electromagnetic contribution  plays a very small role in tunneling considerations, so that this issue is unimportant.

\subsection{The surface tension  \label{sec3-3}}

We now examine the energy density $\sigma$ on the surface of the wall. We ignore the contribution from the negligible magnetic flux on the wall. In flat spacetime, a static wall simply has energy density
\begin{equation}
\sigma = \frac{1}{2 \pi R} \int d \theta \int_{R-\delta/2}^{R+\delta/2} dR \sqrt{h} (f'^2+U) \approx  \int_{R-\delta/2}^{R+\delta/2} dR  (f'^2+U)\,.
\end{equation}
where $h$ is the induced metric on the surface of the wall. For a thin wall $\delta \ll R$ so that the integration is on a relatively small scale. Given this, the Jacobian is approximately constant, $\sqrt{h} \approx R$. Furthermore, a large wall means the equation of motion for the scalar field is approximately
\begin{equation}
f'' = \frac{1}{2}\frac{\partial U}{\partial f} \Rightarrow f'^2 = U - U_{\rm FV} = U\,,
\end{equation}
where $U_{\rm FV}$ corresponds to zero in this paper. Hence, $\sigma$ reduces to
\begin{eqnarray}
\sigma &=& 2 \int^{R+\delta/2}_{R-\delta/2} dR f'^2  = 2 \int_{0}^{1} d f \sqrt{(f^2-\epsilon)(f^2-1)^2} \nonumber \\
&\simeq& 1/2 + \mathcal{O}(\epsilon)\,.
\end{eqnarray}
As the wall expands, there is also a kinetic energy term. Furthermore we must account for a curved spacetime. These two complications are easily treated by working in the Gaussian coordinate frame defined by (\ref{hyperinmetric}). The contribution of the scalar field to the wall Euclidean action will be
\begin{equation}
\int d^3 x_E \sqrt{g} \left[ (\partial_\mu f)^2 + U \right] = 2 \pi \int d \tau_E \tilde{R} \int d \eta \left[(\partial_\eta f)^2 + U \right] = 2\pi \int d \tau_E \, \sigma\tilde{R}\,.
\end{equation}

\section{Decay rate \label{sec4}}

If the scalar field is in a metastable vacuum state, the tunneling process from the false vacuum state to the true vacuum state can occur {\it via} the potential barrier penetration, which is the nucleation process of a vacuum bubble. In the semiclassical approximation, the decay rate of the metastable vacuum state per unit time per unit volume is given by
\begin{equation}
\Gamma/V= A e^{-B}\left[1+ {\cal{O}}(\hbar)\right] \,,
\label{decayrate}
\end{equation}
where the coefficient $A$ comes from the determinant arising in the saddle-point evaluation of the path integral and the exponent $B$ is the difference between the Euclidean action of the bounce solution and the action of the background solution, i.e.\ $B=S_{\text{E, bounce}}-S_{\text{E, bckg}}$. The determinant factor must exclude the integration over the zero modes. The CdL bounce has translation invariance in all direction, giving three zero modes. The vortex only has time translation invariance, giving one zero mode. The position of the vortex is fixed once and for all. These modes are removed. Instead, the corresponding degrees of freedom (either the center of the vacuum bounce in spacetime or the center of the vortex bounce in time) are integrated over. We are interested in finding the exponent $B$.

In presence of a vortex, a similar process exists. As can be deduced in the collective coordinate approximation suggested in Fig.~\ref{fig:eff_pot_a}, quantum fluctuations of the vortex radius can also lead to a phase transition. In this case, the relevant instanton describes the expansion of the wall. To understand the cosmological relevance of the vortex, we will compare the false vacuum lifetime in presence and absence of a vortex. We first proceed to compute the Euclidean action of the relevant instantons in both cases.

\subsection{Ordinary false vacuum decay \label{sec:4-1}}

We imagine the Universe is in a false vacuum state, say $\phi=1$ for definiteness. Ordinary false vacuum decay occurs when a critical true vacuum bubble nucleates and triggers a phase transition. This bubble is the non-trivial extremal path in configuration space which minimizes the Euclidean action, the so-called bounce. The relevant model is a complex scalar field minimally coupled to gravity, whose Euclidean action is given by
\begin{equation}
S_E =\int_{\mathcal{M}} d^3x \sqrt{g} \left [-{\cal{L}}  -\frac{R}{2 \kappa} \right ] - \int_{\partial \mathcal{M}} d^2x \sqrt{h} \frac{K}{\kappa}\,,
\end{equation}
where
\begin{equation}
{\cal{L}} = - (\nabla_{\mu} \phi)^* (\nabla^{\mu} \phi) - U(\phi^* \phi).
\end{equation}

We assume the bounce solution has $O(3)$ symmetry  (as is the case in flat spacetime \cite{Coleman:1977th}), which means $\phi = \phi_B(\rho(\xi))$ where $\xi =\sqrt{\tau^2 + x^2}$ is the Euclidean radial coordinate, $\rho(\xi)$ is the physical radius. We note that this assumption also applies to the trivial solution $\phi = \phi_{\rm FV}$. Also, for simplicity, we assume a real scalar field. With these assumptions, the equation  of motion for the scalar field are
\begin{equation}
\phi'' +\frac{2 \rho'}{\rho}\phi' = \frac{1}{2}\frac{\partial U}{\partial\phi}\,,  \label{eq:EOM}
\end{equation}
where $'$ denotes the differentiation with respect to $\xi$ in this section.  The metric associated with the bounce solution also shares this $O(3)$ symmetry:
\begin{equation}
d s^2 = d \xi^2 + \rho(\xi)^2 d \Omega^2\,.
\end{equation}
$R$ and $K$ are easily found from this metric,
\begin{equation}
R = -\frac{2 \left(2 \rho (\xi ) \rho ''(\xi )+\rho '(\xi )^2-1\right)}{\rho (\xi )^2}\,,
\quad
K = \frac{2 \rho'(\xi)}{\rho(\xi)}\bigg|_{\xi =\infty}\,.
\end{equation}
The action is then given by
\begin{eqnarray}
S_{\rm vac} &=& 4 \pi \int d \xi   \left[ \rho^2(\xi ) (\phi'^2 + U) + \frac{1}{\kappa} \left(2 \rho (\xi ) \rho ''(\xi )+\rho '(\xi )^2-1\right)  \right]- 8 \pi \rho'(\xi) \rho(\xi)\bigg|_{\xi =\infty}\\
&=& 4 \pi \int d\xi   \left[\rho^2(\xi ) (\phi'^2 + U) - \frac{1}{\kappa}\left(\rho '(\xi )^2+1\right) \right]\,,
\end{eqnarray}
where we used integration by parts to cancel the boundary term. The action can be simplified further by using  Einstein's equation $G_{\xi \xi} = \kappa T_{\xi \xi}$ to obtain
\begin{equation}
\rho'(\xi )^2 = 1 + \kappa \rho^2(\xi) \left( \phi'^2 - U\right)\,.
\label{eq:Eins}
\end{equation}
The on-shell action can then be expressed as
\begin{equation}
S_{\rm vac} =  8 \pi \int d \xi \left[ \rho^2(\xi ) U - \frac{1}{\kappa} \right]\,.  \label{eq:Seff}
\end{equation}

Now, the tunneling exponent $B_{\rm vac}$ is obtained by subtracting the background from the action of the bounce
\begin{equation}
B_{\rm vac} \equiv S_{\rm B} - S_{\rm FV} = S_{\rm vac}\bigg|_{\phi = \phi_B(\rho(\xi))} - S_{\rm vac}\bigg|_{\phi = \phi_{\rm FV}}\,.
\end{equation}

We study the specific case where the false vacuum energy density vanishes, $U(\phi_{\rm FV})=0$. $B_{\rm vac}$ then simplifies to
\begin{equation}
B_{\rm vac} = 8 \pi\left[\int_{\rm bounce} d \xi \left( \rho^2(\xi )U-\frac{1}{\kappa}  \right) +  \int_{\rm FV} d \xi  \frac{1}{\kappa} \right]\,.
\end{equation}
We now employ the thin-wall approximation, $\epsilon \ll 1$. In this limit, the bounce solution describes a bubble of true vacuum, centered on the origin $\rho=0$,  which is surrounded by false vacuum. The region of transition from true to false vacuum  is an $O(3)$-spherical wall. Its large radius $\bar{\rho}$ when $\epsilon$ is small explains why its radial profile is dubbed ``thin''. In the thin-wall approximation scheme, the exponent $B_{\rm vac}$ can be divided into three parts: $ B_{\rm vac}=B_{\text{vac, in}} + B_{\text{vac, surface}} + B_{\text{vac, out}}$.

\begin{figure}[ht!]
\begin{center}
\includegraphics[width =4 in]{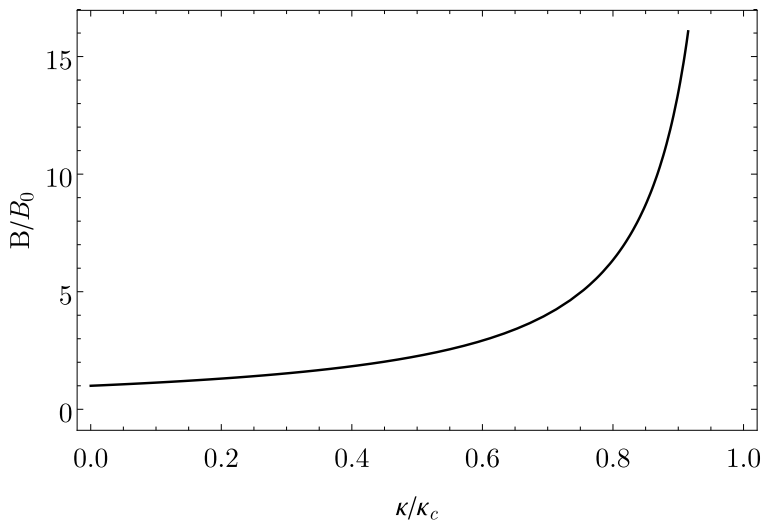}
\end{center}
\caption{The ratio of the tunneling exponent for vacuum decay $B = B_{\rm vac}$ as the function of $\kappa/\kappa_c$.}
\label{fig:B}
\end{figure}

Outside the bubble, the bounce coincides with the false vacuum background, hence $B_{\text{vac, out}}$ vanishes. For the other parts of the spacetime, it is useful to rewrite (\ref{eq:Eins}) as
\begin{equation}
d \xi =  \left[  1 + \kappa \rho^2(\xi) \left( \phi'^2 - U\right) \right]^{-1/2} d \rho \,.
\label{eq:dpdx}
\end{equation}
On the wall, $\rho \gg 1$ and the damping term can be neglected in the equation of motion (\ref{eq:EOM}). We then find the first integral of motion
\begin{equation}
\phi'^2 - U = -U_{\rm FV} = 0\,.
\end{equation}
This means $d \xi = d \rho$ for the bounce in the wall region like in the false vacuum background, and only the potential term contributes to the tunneling exponent. As the wall is thin, the radius doesn't vary much in the wall region, $\rho \approx \bar{\rho}$, and
\begin{equation}
B_{\text{vac, surface}}  = 4 \pi \bar{\rho}^2 \sigma, \quad \sigma \equiv \int_{\rm wall} d \rho \left(\phi'^2 + U \right)\,.
\label{eq:Bwall}
\end{equation}
The on-shell action (\ref{eq:Seff}) inside the bounce is computed with (\ref{eq:dpdx}) and using  $U_{TV} = - \epsilon$, $\phi'=0$,
\begin{eqnarray}
S_{\text{B, in}} &=& - 8 \pi \int_0^{\bar{\rho}} d \rho \left[1+ \epsilon \kappa \rho^2 \right]^{-1/2} \left(\epsilon \rho^2 +\frac{1}{\kappa}\right) \\
&=& -\frac{8 \pi }{\kappa} \left( \frac{1}{2} \bar{\rho} \sqrt{\kappa\epsilon \bar{\rho }^2+1}+\frac{\arcsinh\left(\sqrt{\kappa \epsilon } \bar{\rho }\right)}{2 \sqrt{\kappa \epsilon }}\right) \,.
\end{eqnarray}
As for the inner contribution coming from background, taking $U=U_{\rm FV} = 0$ in (\ref{eq:Seff}) yields
\begin{equation}
S_{\text{FV, in}} = - 8 \pi \int_0^{\bar{\rho}} d\rho \frac{1}{\kappa} = -\frac{8 \pi \bar{\rho}}{\kappa} \,.
\end{equation}
Subtracting this background from the bounce action, we find
\begin{equation}
B_{\text{vac, in}} 	= -\frac{8 \pi}{\kappa} \left(\frac{1}{2} \bar{\rho } \sqrt{\kappa  \epsilon  \bar{\rho }^2+1}+\frac{\arcsinh\left(\sqrt{\kappa \epsilon} \bar{\rho }\right)}{2 \sqrt{\kappa \epsilon} } -  \bar{\rho} \right) \,.
\label{eq:Bin}
\end{equation}
Putting contributions (\ref{eq:Bwall}) and (\ref{eq:Bin}) together, we find the following expression for the tunneling exponent:
\begin{equation}
B_{\rm vac} = 4 \pi \left( \bar{\rho}^2 \sigma +\frac{1}{\kappa} \left[-\bar{\rho } \sqrt{\kappa  \epsilon  \bar{\rho }^2+1}-\frac{\arcsinh\left(\sqrt{\kappa \epsilon } \bar{\rho }\right)}{\sqrt{\kappa \epsilon} } + 2\bar{\rho}
\right] \right) \,.
\label{eq:exp}
\end{equation}
Taking $\kappa \rightarrow  0$ gives the flat spacetime limit, as it must:
\begin{equation}
B_{\rm vac} = \left(4 \pi  \sigma  \bar{\rho }^2-\frac{4}{3} \pi  \epsilon  \bar{\rho }^3\right)+ {\cal{O}} (\kappa) = \left(\sigma {\cal{A}} - \epsilon {\cal{V}}\right)+ {\cal{O}} (\kappa) \,,
\end{equation}
where ${\cal{A}}$ and ${\cal{V}}$ are the flat Euclidean volume and area in three dimensions. The on-shell radius $\bar{\rho}  = \bar{\rho}_0$ found by extremizing $(B_{\rm vac})|_{\kappa=0}$ is
\begin{equation}
\bar{\rho}_0 = \frac{2 \sigma}{\epsilon}. \label{eq:defrho0}
\end{equation}
The tunneling exponent is then evaluated to
\begin{equation}
(B_{\rm vac})\bigg|_{\kappa=0, \;\bar{\rho} = \bar{\rho}_0} \equiv B_0 = \frac{16 \pi \sigma^3}{3 \epsilon^2} \label{eq:defB0}
\end{equation}
The same quantities can be computed in curved spacetime using the full tunneling exponent in (\ref{eq:exp}):
\begin{eqnarray}
\bar{\rho} &=&   \frac{\bar{\rho}_0}{1-\kappa/ \kappa_c}, \quad \kappa_c \equiv \frac{\epsilon}{\sigma^2}, \label{eq:rhostar}\\
 B_{\rm vac} &=& B_0 \left ( \frac{3\left[1 - \frac{(1-\kappa/\kappa_c)}{2 \sqrt{\kappa/\kappa_c}} \arcsinh \left ( \frac{2 \sqrt{\kappa/\kappa_c}}{(1-\kappa/\kappa_c)}\right )\right ]}{2(\kappa/\kappa_c) (1-\kappa/\kappa_c)}\right )= B_0 (1+ {\cal{O}}(\kappa/\kappa_c)) \,.
\label{eq:Bvac}
\end{eqnarray}
We denote by $\kappa_c$ a critical value of $\kappa$. At this value, the bounce radius is infinite, as is the tunneling exponent as shown in Fig.~\ref{fig:B}. Beyond $\kappa_c$, $\bar{\rho}$ would be negative, which is unphysical, so the bounce does not exist for such strong gravity. The vacuum thus becomes completely stable for $\kappa > \kappa_c$. This phenomena, described as gravitational quenching of the vacuum decay, was also observed in $3+1$ dimensions in \cite{Coleman:1980aw}. It is easily seen that $\bar{\rho}$ falls back on $\rho_0$ as gravity is turned off. This is also true for $B_{\rm vac}$ and $B_0$, although it is somewhat less obvious.

\subsection{False vortex disintegration}
According to \cite{Lee:2013ega}, we adopt the ansatz for the configurations representing a vortex of radius $R$, treating $R$ as a variational parameter. For thick-wall vortices
\begin{eqnarray}
f(r) = \left\{
\begin{array}{cc}
r/R & r< R \\
1 & r> R
\end{array} \right. \;, \;\;
a(r) = \left\{
\begin{array}{cc}
(r/R)^2 & r< R \\
1 & r> R
\end{array} \right. \;,
\end{eqnarray}
while for thin-wall vortices
\begin{eqnarray}
  f(r) = \left\{
  \begin{alignedat}{2}
    & 0 &  &  r < R -R_o \\[0ex]
    &\tfrac{r-(R -R_o)}{R_o} & \;\;\; R -R_o <{}&r< R\\[0ex]
    & 1 &		& r>R \\
     \end{alignedat}\right.
 \;, \;\;
a(r) = \left\{
\begin{array}{cc}
(r/R)^2 & r< R \\
1 & r> R
\end{array} \right. \;.
\label{comment1}
\end{eqnarray}

In the following, we focus on the thin-wall solution for which an analytical tunneling exponent can be obtained. The Euclidean action takes the form
\begin{equation}
S_{\rm vort}= \int_{\mathcal M} \sqrt{g}\, d^3 x_E \left[- \frac{R}{2\kappa} + \frac{1}{16\pi}F_{\mu\nu}F^{\mu\nu} + (D_{\mu}\phi)^*(D^{\mu}\phi)+ U(\phi^*\phi)
\right] - \int \sqrt{h} \, d^2x_E \, \frac{K}{\kappa} \,. \label{fvg-eaction}
\end{equation}
The Ricci scalar can be reexpressed by using the trace of the Einstein equations. In three dimensions,
\begin{equation}
-\frac{R}{2 \kappa}=  T =\frac{1}{16\pi} F_{\mu\nu}F^{\mu\nu} - (D_{\mu}\phi)^*(D^{\mu}\phi)-3 U(\phi^*\phi)\, ,
\label{eq:TrEins}
\end{equation}
where we took the trace of (\ref{emten}). Inserting this in (\ref{fvg-eaction}) we obtain the vortex's on-shell Euclidean action
\begin{equation}
S_{\rm vort}= \int_{\mathcal M} \sqrt{g} d^3 x_E \left[\frac{1}{8\pi}F_{\mu\nu}F^{\mu\nu} - 2 U(\phi^*\phi)
\right]  - \int \sqrt{h} \, d^2x_E \, \frac{K}{\kappa} \,, \label{fvg-osaction}
\end{equation}
The final term, the Gibbons-Hawking-York term \cite{York:1972sj,Gibbons:1976ue}, will be written $S_{\rm vort}^{\rm GHY}$ henceforth. It is related to the conical deficit angle as
\begin{equation}
\kappa \mathcal{L}_{\rm vort}^{\rm GHY} = \sqrt{h} K= 1-\Delta = (1- 4 G \mu)^2\,,
\end{equation}
where $\mu$ is the vortex energy. This surface term is not important for tunneling considerations, as it is the same for the background vortex and the expanding vortex.  Indeed, the conical defect is expressed by the energy which is conserved in the Euclidean evolution. Energy considerations are not so obvious once gravity is taken into account, but it makes sense at least for an asymptotic quantity like the surface term. Thus, as we move on and compute the tunneling exponent, the boundary term does not contribute
\begin{equation}
B_{\rm vort}^{\rm GHY} =S^{\rm GHY}_{\text{vort, bounce}} -S^{\rm GHY}_{\text{vort, bckg}} = 0\,.
\end{equation}
Thus, only the bulk contribution of $S_{\rm vort}$ contributes to $B_{\rm vort}$. For simplicity, we write $S^{\rm bulk}_{\rm vort}= S_{\rm vort}$ in what follows. Given the nature of the thin-wall solution, it is better to separate the integral in two parts, that is, the core and the wall of the vortex. The exterior of the vortex does not contribute to the action.
\begin{eqnarray}
S_{\rm vort} &=& \int_{\mathcal M^-} d^3x_E \sqrt{g} \left [ \frac{1}{8 \pi}F_{\mu \nu} F^{\mu \nu} - 2U \right ]
-2 \int_{\Sigma} d^3 x_E \sqrt{g} \,U  \\
&=&  2 \pi \int_{\mathcal M^-} d t_E d r \, r \left [ \frac{1}{8 \pi}F_{\mu \nu} F^{\mu \nu} - 2U \right ]
-4 \pi \int d \tau_E 	\tilde{R} \left (\int_{\Sigma} d \eta \,U  \right )  \\
&=&  2 \pi \int d \tau_E \left [\frac{dt_E}{d\tau_E}\left (\int_0^R d r \, r \left [ \frac{1}{8 \pi}F_{\mu \nu} F^{\mu \nu} - 2U \right ]\right ) - \sigma \tilde{R} \right ] \,.
\end{eqnarray}
For the interior integral, we used the fact that the interior metric respects $\det(g_{\mu \nu}) = r^2$. As for the integral on the wall, we performed it in Gaussian normal coordinates defined by (\ref{hyperinmetric}).
The interior of the vortex is in true vacuum phase,  $U=-\epsilon$, so
\begin{equation}
\int_0^R d r \, r U =  - \frac{\epsilon R^2}{2}\,.
\end{equation}
Furthermore, the bounce solution is determined by (\ref{junceq}), as the radius expands between turning points $R_0$ and $R_1$ as shown in Fig.\ref{fig:eff_pot_a}. The integration over Euclidean time can be parameterized with the radius $R(\tau_E)$.
\begin{equation}
S_{\rm vort} = 2\pi \int_{R_0}^{R_1} d R \frac{1}{\dot{R}} \left[ \frac{dt_E}{d\tau_E}  \left( \int_0^R d r \, r \frac{1}{8 \pi} F_{\mu \nu} F^{\mu \nu}+\epsilon R^2 \right) - \sigma \tilde{R} \right]\,,
\end{equation}
where $\dot{R}=dR/d\tau_E$. The proper time is related to the coordinate time through
\begin{equation}
d \tau_E^2  = g_{00} d t_E^2+ g_{11} d \tilde{R}^2 \Rightarrow
\frac{dt_E}{d\tau_E}	
= \left[ \frac{1 - g_{11} \dot{R}^2 \left(\frac{d\tilde{R}}{dR}\right)^2 }{g_{00}} \right]^{1/2}
= \left[ \frac{1 - \dot{R}^2 \left( \frac{H}{N^2 J^2} \right)}{N^2} \right]^{1/2}\,.
\end{equation}
The action then reads
\begin{equation}
S_{\rm vort} = 2 \pi \int_{R_0}^{R_1} d R \frac{1}{\dot{R}} \left[ \left[ \frac{1 - \dot{R}^2 \left( \frac{H}{N^2 J^2} \right)}{N^2} \right]^{1/2} \left( \int_0^R d r \,  r \frac{1}{8 \pi} F_{\mu \nu} F^{\mu \nu}+\epsilon R^2 \right)
- \sigma R \sqrt{J/L} \right]\,.  \label{eq:SEdev}
\end{equation}

To evaluate $F_{\mu \nu} F^{\mu \nu}$, we first compute the value of the gauge field with a computation analogous to the definition of the magnetic flux calculated in eq.\ (\ref{totalflux})
\begin{equation}
\frac{1}{2\pi} \oint A_\mu dx^\mu = A_\theta = \frac{1}{2 \pi} \int^{r}_{0} \int^{2\pi}_{0} d^2 x \sqrt{\det g_{ij}}{\mathcal B} =  \frac{Q_m}{2 L} \ln (1+L \Lambda r^2) \,.
\end{equation}
The radial integral of the field-strength term then becomes
\begin{eqnarray}
&&\int_0^R d r\,  r \frac{F^2_{\mu \nu}}{8 \pi}
= \int_0^R d r \frac{r}{4 \pi} \left( g^{tt} g^{\theta \theta} \left (\partial_t A_\theta \right )^2 + g^{rr} g^{ \theta \theta} \left (\partial_r A_\theta \right )^2  \right) \\ \nonumber
&& = \int_0^R dr\,r \left (
\frac{\pi \dot{R}^2 \left(Q_m'\right)^2 \left(\log (N^2) \left[-\frac{\kappa^2 \Lambda ^3 r^2 Q_m^4}{64 \pi^2}-\frac{\kappa \Lambda  Q_m^2}{8 \pi}+\Lambda  r^2+1\right]+ \frac{\kappa \Lambda ^2 L r^2 Q_m^2}{4 \pi}\right)^2}{16 L^2 N^6 \left( \frac{\kappa Q^2_m}{8 \pi} \log (N^2)+L r^2\right)}
+ \frac{\Lambda ^2 Q_m^2}{4 \pi  N^2}\right), \nonumber
\label{eq:intF2}
\end{eqnarray}
where the time dependence originated from $Q_m$ which is a function of $R(t)$. This integral is quite complicated and cannot be done analytically. Fortunately, we can simply neglect this contribution from the action. The general idea is that the action is of the form
\begin{equation}
S_{\rm vort} = \int^{R_1}_{R_0} d R \, \left ( \text{function of} \; R\right ) \,.
\end{equation}
Since $R_1 \sim 1/\epsilon$ is very large (and gets bigger as gravity is added), only the highest powers of $R$ in the integrand will have significant contributions in the large $R$ part of the integration. Terms with smaller powers of $R$ will be negligible. We also take into account that $\epsilon R^2 \sim R$ for $R \sim R_1$. Based on this criterion, it was argued in \cite{Lee:2013ega} that the vortex action (in flat spacetime)
is
\begin{eqnarray*}
&&\int^{R_1}_{R_0} dR  \left [\int dr \,r \frac{1}{16 \pi} F_{\mu \nu} F^{\mu \nu} + |D_\mu \phi|^2 +U \right]\\ &&\qquad\qquad\qquad\qquad= \int d R \left [ \frac{\Phi^2}{8 \pi^2  R^2} \left (1+\frac{\dot{R}^2}{2} \right ) + 2 \pi \sigma R \sqrt{1+\dot{R}^2} -  \pi \epsilon R^2 \right ]\\
&&\qquad\qquad\qquad\qquad\approx \int d R \left [ 2 \pi \sigma R \sqrt{1+\dot{R}^2} -  \pi \epsilon R^2 \right ] \,.
\end{eqnarray*}
We should verify if this approximation is still valid in the presence of gravitational corrections. For simplicity, we restrict our demonstration to the static contribution to the curved spacetime integration of the electromagnetic field strength. We expand this expression in powers of $\kappa$:
\begin{eqnarray}
\int_0^R d r \, 2	r  \left ( F_{12} F^{12} \right ) &=& \int_0^R d r \, 2 r \frac{\Lambda^2 Q_m^2}{N^2} \nonumber \\
 &=&  \int_0^R d r \,  2 r \left(\frac{\Phi ^2}{\pi ^2 R^4} + \frac{\Phi ^2 \left(R^2-r^2\right) \left(8 \pi ^3 R^4 \epsilon +\Phi ^2\right)}{8\pi ^5 R^8} \kappa + {\cal{O}}(\kappa^2)\right) \nonumber \\
&=&
\frac{\Phi ^2}{\pi ^2 R^2}+\frac{\Phi ^2  \left(8 \pi ^3 R^4 \epsilon +\Phi ^2\right)}{16 \pi ^5
 R^4} \kappa  + \mathcal{O}(\kappa^2) \,,
\end{eqnarray}
where in the second step we have rewritten $Q_m$ in terms of $\Phi$ by solving \eqref{totalflux} to lowest order in $\kappa$.
All these terms are either inverse powers of $R$ or are constant with respect to $R$ and will contribute only weakly to the value of the action. To fully justify this approximation, we would also need to determine gravitational corrections for the kinetic analog of this quantity, and then compare them to the scalar field contribution. We will first compute the action completely ignoring the electromagnetic part. We will later verify our approximation with a numerical, non-perturbative, computation.

With all this, $B_{\rm vort}$ is reduced to
\begin{equation}
B_{\rm vort} \approx 2 \pi \int_{R_0}^{R_1} d R \, \frac{1}{\dot{R}} \left[ \left[ \frac{1 - \dot{R}^2 \left( \frac{H}{N^2 J^2} \right)}{N^2} \right]^{1/2} \!\!\!\! \epsilon R^2 - \sigma  R \sqrt{J/L} \right] \,.
\end{equation}

We apply the same approximation, keeping only highest powers of $R$ and using $\epsilon R^2 \sim R$. The effective potential (\ref{efpotential}) related to the wall velocity $\dot{R}$  is simplified to
\begin{equation}
\dot{R}^2 = 2 V_{\rm eff}(R) \approx  \left (1 - R^2\frac{(1 - \kappa/\kappa_c)^2}{4 \sigma^2/\epsilon^2} \right ) =  \left (1 - R^2/R_1^2 \right ), \quad R_1 \equiv \frac{2 \sigma / \epsilon}{|1-\kappa/\kappa_c|} \,.
\label{eq:Rdotapprox}
\end{equation}
Note that for $\kappa < \kappa_c$, the turning point $R_1$ is exactly the radius of the CdL bounce defined in (\ref{eq:rhostar}). It is convenient to introduce a variable for the ratio $R/R_1$, such that
\begin{equation}
\dot{R} = \sqrt{1-R^2/R_1^2} = \sqrt{1-x^2}, \quad x \equiv R/R_1 \,.
\end{equation}
Finally, approximating $Q_m$ to $0$ also yields $J \approx H\approx L \approx 1$.  In its final form, the tunneling exponent then reads
\begin{eqnarray}
B_{\rm vort} &\approx& 2 \pi \sigma \int_{R_0}^{R_1} d R \frac{1}{\dot{R}} \left[ \frac{1}{N^2} \left[ N^2 - \dot{R}^2 \right]^{1/2} \frac{\epsilon}{\sigma} R^2 - R \right]\\
&=& 2 \pi \sigma R_1^2 \int_{0}^{1} d x \frac{1}{\sqrt{1-x^2}} \left[ \frac{(\epsilon/\sigma)R_1 x^3}{1+\kappa \epsilon R_1^2 x^2} \sqrt{1 + \kappa \epsilon R_1^2} - x \right] \,,
\label{eq:preBvort}
\end{eqnarray}
where we used $R_0 \ll R_1$ to set $x \in [0,1]$, and we used $N^2 = 1 + \kappa \epsilon R^2$ from (\ref{eq:defs}). Replacing $R_1$ by its value defined in (\ref{eq:Rdotapprox}), we get
\begin{equation}
B_{\rm vort} = \frac{8 \pi  \sigma ^3 /\epsilon^2 }{\left(1-\kappa/\kappa_c\right)}\int_0^1 dx \, \frac{x \left(2 x^2+ \kappa/\kappa_c-1\right)}{ \left ( \left(1-\kappa/\kappa_c\right)^2+4 \kappa/\kappa_c \, x^2\right ) \sqrt{1-x^2} } \,.
\end{equation}
Proceeding with the integration and using $B_0$ defined in (\ref{eq:defrho0}) to simplify the prefactor, we obtain
\begin{equation}
B_{\rm vort} = \frac{B_0}{2} \left(\frac{3 \left(1-\frac{(1-\kappa/\kappa_c)}{2 \sqrt{\kappa/\kappa_c}} \arctanh \left(\frac{2 \sqrt{\kappa/\kappa_c}}{1+\kappa/\kappa_c}\right)\right)}{2(\kappa/\kappa_c) (1-\kappa/\kappa_c)} \right) \,.
\label{eq:Bvort}
\end{equation}
Since
\begin{equation}
\arctanh
\left ( \frac{2 \sqrt{\kappa/\kappa_c}}{1+\kappa/\kappa_c}\right ) =
\arcsinh \left ( \frac{2 \sqrt{\kappa/\kappa_c}}{|1-\kappa/\kappa_c|}\right ) \,,
\end{equation}
we conclude from \eqref{eq:Bvac} that
\begin{equation}
B_{\rm vort}  =  B_{\rm vac}/2, \quad \kappa < \kappa_c \,.
\end{equation}
Thus, the vortex and vacuum decay rates share the same dependence on $\kappa$ for $\kappa < \kappa_c$. The comparison does not hold for $\kappa > \kappa_c$. As we mentioned in Sec.~\ref{sec:4-1}, regular vacuum decay is forbidden in this region as the radius of the bounce becomes negative. No such restriction applies to the vortex; as $\kappa$ becomes greater than $\kappa_c$, Fig.~\ref{fig:eff_pot_a} and (\ref{eq:Rdotapprox}) show that the false vortex's escape radius $R_1$ remains positive. However, we do note a singular behaviour near $\kappa_c$:
\begin{equation}
\lim_{\kappa \to \kappa_c^{\pm}} B_{\rm vort} \to \mp \infty \,.
\end{equation}
The lower case gives an effect similar to the quenching of the CdL bounce as noted earlier: a change of sign happens as $\kappa$ goes from $\kappa_c^- \to \kappa_c^+$.

As $\kappa$ increases, this analytical result may not hold, since $R_1$ gets smaller and we may no longer neglect a portion of the integration on the interval $[R_0, R_1]$. We must also emphasize that to even consider a tunneling exponent, a metastable vortex must exist in the first place. As was pointed out in Section \ref{sec2-2}, there is a maximum value of $\kappa$ beyond which static solutions of the equations of motion cannot be found. From Fig.~\ref{fig:thin}, we see a vortex solution corresponding to $\kappa = 0.04$ respects $\kappa > \kappa_c$ since $\kappa_c = \epsilon / \sigma^2 \approx (0.005)/ (1/2)^2 = 0.02$. This means there is a portion in region $\kappa > \kappa_c$ where the CdL bounce becomes impossible, while the formation and decay of false vortices is still possible. In fact, the vortices in question are very short-lived since $-B_{\rm vort} \gg 1$. The thought of a gravitationally stabilized false vacuum may have given us hope in the past; with the prospect of gravitationally enhanced vortex explosions, we  have occasion for renewed anxiety, to put in Coleman's words \cite{Coleman:1977py}.

\begin{figure}[ht!]
\begin{center}
\includegraphics[width =3.2 in]{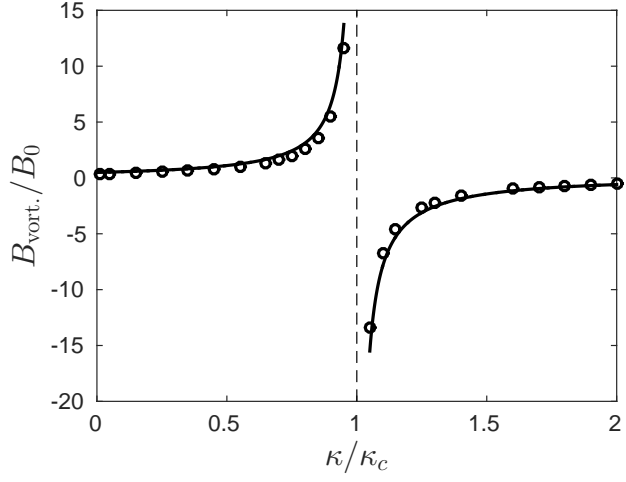}
\end{center}
\caption{Full line: analytical approximation of $B_{\rm vort}/B_0$ as a function of $\kappa/\kappa_c$, \eqref{eq:Bvort}. Circles: Numerical solution with $Q_m \neq 0$ ($\Phi =100$ and $\epsilon=0.01$).}
\label{fig:verif}
\end{figure}

We numerically verify the approximation scheme used to obtain (\ref{eq:Bvort}). The numerical value of the action is obtained by inserting (\ref{eq:defs}, \ref{backcosyre}, \ref{junceq}, \ref{eq:intF2}) into the full bounce action (\ref{eq:SEdev}). Parameters $Q_m$ and $\Delta$ must also be replaced by minimizing (\ref{totalflux}) and solving (\ref{deltaconide}), respectively. The result of the integration  is shown in Fig.\ref{fig:verif};  there is excellent agreement between the analytical and numerical computations.

\begin{figure}[h!]
\begin{center}
\includegraphics[width = 0.65 \linewidth]{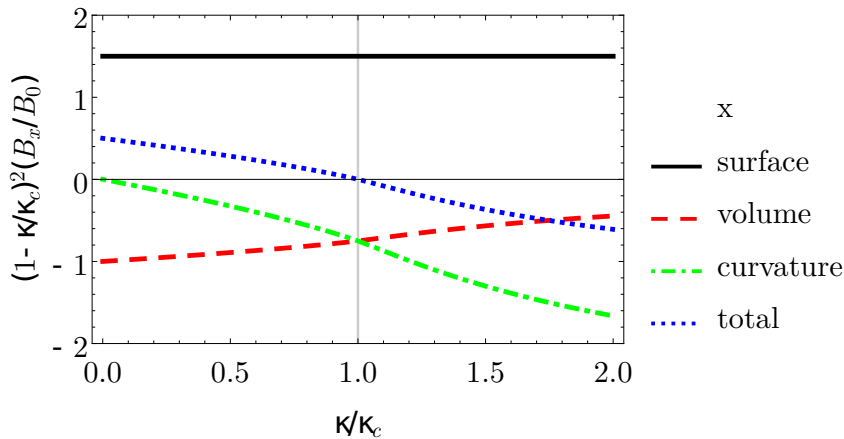}
\end{center}
\caption{Contributions $B_x$ as a function of $\kappa/\kappa_c$ to the tunneling exponent. The line patterns (solid line, dashed line, dash-dotted line, dotted line) respectively represent $B_{\rm surface}$, $B_{\rm volume}$, $B_{\rm curvature}$ and $B_{\rm total}$.}
\label{fig:Bx_B0}
\end{figure}

Relevant contributions to the tunneling exponent are given by
\begin{equation}
B_{\rm surface} + B_{\rm volume} + B_{\rm curvature} = B_{\rm total} =  B_{\rm vort} \,,
\end{equation}
where
\begin{eqnarray}
B_{\rm surface} &\equiv& \int_{\Sigma} d^3x_E \sqrt{g} \left ( \phi'^2 + U(|\phi|) \right ) = \int_{\Sigma} d^3x \sqrt{g} \left ( 2 U(|\phi|) \right )\,,\\
B_{\rm volume} &\equiv&  \int_{\man^-} d^3x \sqrt{g} \left (U(|\phi|) \right )\,,\\
B_{\rm curvature} &\equiv& \int_{\man} d^3x \sqrt{g} \left (-\frac{R}{2 \kappa} \right )\\ \nonumber
&=& \int_{\man} d^3x \sqrt{g} \left (-\phi'^2 - 3 U(|\phi|) \right ) = - 2 B_{\rm surface} - 3  B_{\rm volume}\,,
\end{eqnarray}
where again we used (\ref{eq:TrEins}) to reexpress the Ricci curvature. We already computed a combination of these terms to obtain (\ref{eq:Bvort}). We repeat the procedure for each individual term:
\begin{eqnarray}
B_{\rm surface}/B_0 &=& \frac{3}{2 (1-\kappa/\kappa_c)^2}\,,\\
B_{\rm volume}/B_0 &=& -\frac{6 \left(1+\kappa/\kappa_c\right) \sqrt{\kappa/\kappa_c}-3 \left(1-\kappa/\kappa_c\right){}^2 \tanh ^{-1}\left(\frac{2 \sqrt{\kappa/\kappa_c}}{1+\kappa/\kappa_c}\right)}{16 \left(1-\kappa/\kappa_c\right){}^2 \left(\kappa/\kappa_c\right){}^{3/2}}\\ \nonumber
&\approx& -\frac{3 \left(1+\kappa/\kappa_c\right)}{8 \left(1-\kappa/\kappa_c\right){}^2 \kappa/\kappa_c}, \quad \kappa \to \kappa_c\\ \nonumber
&=& \left (\frac{1+\kappa/\kappa_c}{4 \, \kappa/\kappa_c} \right ) B_{\rm surface}/B_0\\
B_{\rm curvature}/B_0 &=& - \left (2 + \frac{3}{4}\, \frac{1+\kappa/\kappa_c}{\kappa/\kappa_c}\right )  B_{\rm surface}/B_0 \,.
\end{eqnarray}
These terms are of the form $B_x / B_0$ with $x \in \{\text{surface, volume, curvature} \}$. It is easier  to compare $(1- \kappa/\kappa_c)^ 2 \left (B_x / B_0 \right )$ since the singular part is then gone. Fig.~\ref{fig:Bx_B0} shows how these terms, as well as their sum, change as $\kappa / \kappa_c$ is varied. Not surprisingly, we see that the curvature part of the action is mostly responsible for the behavior and the change of sign of the tunneling exponent $B_{\rm total} = B_{\rm vort}$. The change of sign, which occurs  for $\kappa / \kappa_c$, is a result of a simplification
\begin{equation}
\lim_{\kappa \to \kappa_c}\sum_x B_x \sim \frac{1-\kappa/\kappa_c}{(1-\kappa/\kappa_c)^2} = \frac{1}{1-\kappa/\kappa_c}\,.
\label{eq:Bvort_sign}
\end{equation}

It may seem surprising that the vortex tunneling exponent \eqref{eq:Bvort} is independent of the magnetic flux $\Phi$. This is because this quantity only comes into play in the contributions we argued were negligibly small. Put another way, the vortex disintegration is essentially controlled by the wall surface tension and the vortex vacuum volume energy. In our approximation scheme, we really just computed the tunneling exponent of an $O(2)$-symmetric bubble which starts from approximately null radius and grows to the escape radius in Euclidean time. Now, this $O(2)$-invariant tunneling event with lower action than the $O(3)$-invariant tunneling event does not contradict what we know from regular vacuum decay.  In the absence of magnetic flux, the former is not a proper decay channel since it does not extremize the action. By breaking translational invariance, the vortex basically enables this mode. The magnetic flux is thus necessary for the existence of this event, but has a minor influence on its occurrence. Apart from the tunneling exponents which vary by a factor $2$, the major difference between the $O(2)$ and $O(3)$ bounces is that only the former is still well-defined for $\kappa > \kappa_c$.

\subsection{Tunneling rates}
We compare decay rates  for $\kappa < \kappa_c$ where both vortex and CdL bounces are possible. For a dilute gas of instantons, the decay rate in the semi-classical approximation is given by (\ref{decayrate}). For the coefficient $A$, the change of variables gives rise to a Jacobian factor which is evaluated in \cite{Coleman:1977py} and yields the decay rate
\begin{equation}
\Gamma =A' L^{(\# \text{zero modes} - 1)}  \left(\frac{B}{2\pi}\right)^{(\# \text{zero modes})/2}  e^{-B} \,,
\end{equation}
where $A'$ is the determinant excluding the zero mode and $L$ denotes dimensions of space or time. We compare the decay rate for vortex disintegration and regular vacuum decay. The vortex tunneling rate has to be multiplied by the number $\mathcal{N}$ of vortices. It is assumed that vortices are sufficiently separated such that intervortex interactions can be ignored. We thus write tunneling rates ratio
\begin{equation}
\frac{\Gamma^{\rm vac}}{\mathcal{N}\Gamma^{\rm vort}} = \frac{V A'^{\rm vac}\left(\frac{B_{\rm vac}}{2\pi}\right)^{3/2}\, \exp\{-B_{\rm vac}\}}{\mathcal{N} A'^{\rm vort}\left(\frac{B_{\rm vort}}{2\pi}\right)^{1/2}\, \exp\{-B_{\rm vort}\}}
= \frac{ A'^{\rm vac} } {(\mathcal{N}/V) A'^{\rm vort} } \frac{ \sqrt{2} B_{\rm vac}}{2 \pi} \exp \left  \{-\frac{B_{\rm vac}}{2}\right \}\,,
\end{equation}
where we used $B_{\rm vort} = B_{\rm vac}/2$. $\mathcal{N}/V$ indicates the vortices density. Of course, let us recall that we assumed  from the outset that $\epsilon \ll 1$ and that the vortex has a large winding number $n$. Since $B_{\rm vac}$ is very large, and more so as $\epsilon \to 0$ and/or $\kappa \to \kappa_c$, this means the phase transition is largely dominated by vortex disintegration. Calculating the determinant factors $A'^{\rm vac}$ and  $A'^{\rm vort}$.  is beyond the scope of this paper

\section{Summary and Discussion \label{sec5}}

We have extended the work \cite{Lee:2013ega} based on a modified Abelian Higgs model in which vortices can be formed in a $U(1)$ breaking false vacuum. We found how gravitational effects can alter the formation and decay rate of vortices trapped in the false vacuum. As gravity is turned on, the spacetime becomes asymptotically conical, with a deficit angle clearly seen in numerical solutions. Matter configurations are also changed, albeit to a lesser degree. As for the decay rate, it decreases, both for conventional vacuum decay and for vortex disintegration, as $\kappa$ is increased towards its critical value $\kappa_c$. Neglecting the magnetic contribution in the vortex case, we find the vortex tunneling exponent is precisely half that for vacuum decay.  Increasing $\kappa$ up to this critical point $\kappa_c$, both these events become more and more suppressed. Beyond this critical point, the CdL bounce is completely suppressed, while the vortex bounce becomes extremely favored.  However, $B_{\rm vort}$ remains a monotonic increasing function of $\kappa$.

We compared tunneling amplitudes in the region $\kappa < \kappa_c$. We found out that, as gravity is turned on, vortices remains the dominant factor determining the false vacuum's stability. As in flat spacetime, the overall decay rate and the vortex's dominance increase as $\epsilon \to 0$. Thus, in some cases, vortices may very well render an otherwise-acceptable theory incompatible with our Universe's cosmological history. We should stress out that gravity in the $\kappa < \kappa_c$ region stabilizes solutions of the model by decreasing the tunneling decay rate. It also increases the region of parameter space for which there are classically stable vortices, since the vacuum energy density limiting value, $\epsilon \rightarrow \epsilon_c$, is increased as gravity is turned on.

Of course, the model in question has to appear in physical situations in the first place. One of our motivations was to study the interplay of symmetry breaking and false vacuum in a toy model, in which we have presented the numerical solutions and analytical calculations. These features are of general relevance in many theories. One can think of scalar potential false vacua appearing in string cosmology, or the existence of supersymmetry-broken phases. More complete but similar models to ours for Grand Unified Theories were also considered at a qualitative level in \cite{Steinhardt:1981ec}.

Vortices have been studied in a variety of theories \cite{Hong:1990yh, Jackiw:1990aw, Lee:1990it, Lee:1992yc, Cho:2007jf, Kim:2009ny}. Especially in models with the sextic potential, we expect that the models can open up the possibility that various types of solutions exist both with and without gravity. Decaying cosmic strings, which generalize the false vortex in $3+1$ dimensions, were also studied in earlier work \cite{Lee:2013zca}. For the classical instability the effect was first investigated in the context of Grand Unified Theories in \cite{Yajnik:1986tg}. In those studies, cosmic strings are analogous to vortex lines in type II superconductors or in superfluid liquid helium. The extension to the decay of those with gravity could be interesting.

Different topological solitons such as  metastable monopoles \cite{Kumar:2010mv} and domain walls \cite{Dupuis:2015fza, Haberichter:2015xga} were also studied in similar models with flat spacetime. Obtaining a generalization in curved spacetime for these defects would also be interesting. For example, early results in the monopole case show the same sign-changing of the tunneling exponent noted in \eqref{eq:Bvort_sign}. More investigations along these lines are under way.

\section*{Acknowledgements}

We thank Cliff Burgess and Misao Sasaki for useful discussions. \'ED thanks Marie-Lou Gendron Marsolais for her help with numerical computations.  This work was financially supported in part by the Natural Sciences and Engineering Research Council of Canada and the Fonds de recherche du Qu\'ebec --- Nature et technologies. RM, MP and UY thank the Ministère des Relations internationales et de la Francophonie du Gouvernement du Qu\'ebec under the Cooperation Qu\'ebec-Maharashtra for continuing financial support. BHL is supported by Basic Science Research Program through the National Research Foundation of Korea (NRF-2017R1D1A1B03028310). WL is supported by Basic Science Research Program through the National Research Foundation of Korea (2016R1D1A1B01010234).  DY is supported by the Leung Center for Cosmology and Particle Astrophysics (LeCosPA) of National Taiwan University (103R4000) and supported from the Korea Ministry of Education, Science and Technology, Gyeongsangbuk-Do and Pohang City for Independent Junior Research Groups at the Asia Pacific Center for Theoretical Physics.

\end{document}